# Analysis of User Experience Evaluation Methods for Deaf users: A Case Study on a mobile App

Andrés Eduardo Fuentes-Cortázar[1,2] [0000-0002-4620-6106], Alejandra Rivera-Hernández[1,3] [0009-0004-8739-8601] and José Rafael Rojano-Cáceres* [1,4] [0000-0002-3878-4571]

[1] Faculty of Statistics and Informatics, University of Veracruz, Xalapa, Veracruz, Mexico
[2] `afuentescortazar@gmail.com`, [3] `alejandra.rihe99@gmail.com`
[4] `rrojano@uv.mx`

**Abstract.** User Experience (UX) evaluation methods that are commonly used with hearing users may not be functional or effective for Deaf users. This is because these methods are primarily designed for users with hearing abilities, which can create limitations in the interaction, perception, and understanding of the methods for Deaf individuals. Furthermore, traditional UX evaluation approaches often fail to address the unique accessibility needs of Deaf users, resulting in an incomplete or biased assessment of their user experience. This research focused on analyzing a set of UX evaluation methods recommended for use with Deaf users, with the aim of validating the accessibility of each method through findings and limitations. The results indicate that, although these evaluation methods presented here are commonly recommended in the literature for use with Deaf users, they present various limitations that must be addressed in order to better adapt to the communication skills specific to the Deaf community. This research concludes that evaluation methods must be adapted to ensure accessible software evaluation for Deaf individuals, enabling the collection of data that accurately reflects their experiences and needs.

## 1  Introduction

User experience (UX) evaluation is an essential aspect of digital application design, as it aims to ensure that interfaces are intuitive, easy to use, and accessible to all users, regardless of their individual characteristics. UX has been extensively documented in various research studies, leading to a variety of proposals in the form of tools, methods, and evaluation techniques [18]. This phenomenon has resulted in multiple interpretations of the UX concept [19],[20],[21]. The execution of UX testing has played a crucial role in the software development lifecycle, as without these evaluation methods, it is impossible to determine whether products meet the needs and expectations of users [22]. Moreover, through this practice, a deep understanding of target users is gained, encompassing both their behavior when interacting with the system and their impressions of the experience lived [7]. However, the methods and tools used to evaluate UX are based on a wide range of assumptions, leading to uncertainty among researchers regarding which methods are most appropriate for assessing the different aspects of the user experience [2].

As technology advances, the importance of an inclusive user experience (IUX) becomes increasingly evident, as more and more individuals with diverse abilities or disabilities engage with applications in their daily lives [15]. User-centered design (UCD) emphasizes the active participation of users from the early stages of system development, aiming to create applications that genuinely meet their needs [8]. In this context, when developing software applications for individuals with hearing disabilities, it is crucial for developers and designers to understand the characteristics and specificities of the Deaf community. Additionally, they must engage potential users in all phases of the process and conduct continuous testing throughout development to ensure that the fundamental factors proposed by user experience are met.

There are various studies focused on software development for Deaf individuals; however, many of these do not adequately address the implementation of accessible UX evaluation methods. Traditional UX evaluation methods, commonly used with hearing users, may not be functional or effective when applied to Deaf users. These methods, primarily designed for individuals with auditory capabilities, can create barriers in interaction, perception, and understanding of applications for Deaf users [10].

In relation to accessibility during testing with people with disabilities (PWD), it becomes evident that the evaluation of software applications for Deaf individuals requires methods and procedures specifically adapted to their needs. This is primarily due to the particular communication and linguistic needs of sign language users [9]. One of the implications associated with hearing loss is the difficulty in accessing written language and the challenge of developing adequate and functional reading skills [6]. Furthermore, the literacy limitations of Deaf individuals may hinder their understanding of content that is not accessible [14].

The primary objective of this study is to evaluate empirically a range of UX assessment methods commonly used by HCI researchers with Deaf users, with the aim of determining their accessibility, effectiveness, and the quality of the results they yield. By testing these methods in real-world conditions, the study seeks to understand how well they adapt to the needs of Deaf individuals, providing insights into which approaches are most suitable for evaluating user experience in this specific context. Furthermore, the study aims to identify any potential challenges or barriers faced by Deaf users during these assessments, ultimately contributing to the improvement of UX evaluation practices for this community.

## 2    Methodology of experimentation

The methodological approach adopted in this research is empirical in nature, as it prioritizes the direct observation, participation, and interaction with users as the primary source of knowledge construction [16]. For us, this is particularly relevant when working with populations with specific accessibility needs such as Deaf users, since their experience, language, and interaction with technological systems often diverge from the standard assumptions, which are found in traditional usability frameworks.

The decision to structure the research in phases responds to both practical and methodological considerations to ensure the organized and effective development of the study. Therefore, we considered the specific objectives of the study, the particular characteristics of the participant population, and grounded in the authors' prior experience with related research. As a result, the following stages were defined:

For the execution of this research, a series of stages were established, which were essential to ensure the organized and effective development of the study. These stages were carefully considered based on the objectives of the work.

1. **Selection of the Population**: In this phase, the target population for the study was identified and defined, focusing on Deaf users. This selection was aligned with the objective of the investigation, which focused on UX evaluation from an accessibility-oriented perspective, taking into account the communicative and cognitive particularities of this user group.
2. **Formation of the Evaluation Team**: In this phase, an evaluation team was formed by three researchers and a mediator. The UX researchers served as evaluators with experience in assessments involving Deaf users, and the mediator is the teacher's group. In addition, the teacher is our pedagogical expert and also, she played a crucial role as an interpreter between the technical team and the participants. Her presence was essential for ensuring effective communication, as well as providing accurate cultural and linguistic interpretation throughout the entire process.
3. **Selection of UX Evaluation Methods**: In this phase, the UX evaluation methods specifically found in the literature for use with Deaf users were selected. The selection was based on a previous literature review, as well as prior recommendations from similar studies. In addition, the suitability of the methods was assessed in relation to the children's age, cognitive abilities, and level of participation. The opinion of the teacher-mediator was crucial in determining the selection of each method within the specific context of the group.
4. **Planning of Tests and Activities**: Based on the selected evaluation methods, a detailed plan was developed for the tests and activities to be carried out throughout the research. This planning included the sequence of tasks, the necessary materials, estimated execution times, and support strategies. The goal was to create an accessible and engaging environment that would encourage active participant interaction and allow the collection of meaningful information about their experience with the app.
5. **Execution of Evaluations and Data Collection**: Finally, the evaluations were carried out in a controlled environment, with the constant presence of the research team and the teacher. During this phase, users' interactions with the app were carefully observed, their behaviors, reactions, and comments were recorded, and qualitative data were collected for subsequent analysis.

Once the various phases that structure this research have been defined, the subsequent sections will delve into the analysis of the implementation of different UX evaluation methods with Deaf users.

## 3      Methods for evaluating UX focused on Deaf users

The evaluation techniques or methods commonly used with hearing individuals may not be effective for Deaf individuals due to the significant difference in their corresponding natural language: oral language vs sign language. Therefore, it is important to choose evaluation instruments that are accessible according to the characteristics of Deaf individuals. Conducting accessible UX tests ensures that applications used by Deaf users meet their needs and provide a positive experience [1].

As stated, in a previous study, a literature review was conducted to identify UX evaluation methods and their application to users with hearing difficulties or Deaf individuals [3]. The following are the UX evaluation methods suggested for using with Deaf individuals in software testing: a) Think aloud protocol, b) Drawing intervention, c) Picture card, d) Wizard of Oz, e) Surveys, f) Smileyometer, g) Interviews, h) Focus group, i) Cognitive walkthrough, j) Direct observation, k) Fun toolkit, l) EMODIANA, m) User workshop.

For each case proposed, it was also identified some adaptations such as the inclusion of Sign Language, the use of simple scales, and ensuring that the tools do not rely on verbal information stand out.

Once the different accessible UX evaluation methods were identified, the next step was to select those that would be part of a set of tests with Deaf users. For this purpose, each method was presented to the teacher's group.

The selection of methods was based on her feedback and suggestions, taking into account factors such as the children's abilities, the level of interactivity of the techniques, with the aim of encouraging active participation, as well as their prior use in other evaluation studies.

As a result, the following UX evaluation methods were considered: a) User workshop, b) Cognitive walkthrough, c) Direct observation, d) Picture card, and e) EMODIANA. In the following section a brief explanation is presented for each one.

### 3.1      UX evaluation methods selected to be applied

The following UX evaluation methods were selected for use with Deaf children, chosen for their ability to accommodate the participants' specific needs and ensure inclusive and effective assessments. These methods were implemented across multiple planned sessions, which are described in section 4.2.

**User workshop**
This participatory design method brings users and designers together in focused sessions [5].

**Direct observation**
Direct observation is a completely observational study that provides a more authentic view of practices and needs from users [13].

**Cognitive walkthrough**

A cognitive walkthrough evaluates how easily new users complete tasks, noting time and errors [12].

**Picture card**

This method uses illustrated and customizable cards for participants to share and draw experience-based stories [5].

**EMODIANA**

It measures emotions subjectively using colored characters that represent ten distinct emotional states [4].

## 4     Case study

This study was conducted at the Multiple Attention Center (CAM by its acronym in Spanish) in a city in Mexico. This school provides specialized education to children and young people living with disabilities, conditions that hinder their enrollment in regular schools. At the CAM, the educational practice of its professionals is framed within the current Education Plan and Curriculum Programs in Mexico [11]. In a classroom, there can be students who live with different types of disabilities.

For the purposes of this research, participants are divided into two groups, allowing us to differentiate the results obtained from the app tests. **Table 1** we describe all the participants of both groups. It should be said that parental consent was obtained, including the only girl which belongs to high school, the rest of students belong to elementary school.

Sign Language is taught to all students (Group A and Group B), addressing its syntax, grammar, and vocabulary in an integrated manner. In the classroom, a wide variety of visual teaching materials are presented, specifically designed to support the learning of deaf students. This material includes visual representations of the alphabet, days of the week, months of the year, school supplies, animal names, among others. Each element is presented clearly to help students associate signs with everyday concepts, while progressively familiarizing them with the structure and vocabulary of Sign Language, reinforcing their ability to communicate in daily situations.

Table 1. Demography for user group.

| Group | Members | Gender | Age |
|---|---|---|---|
| (A) Deaf users | Profoundly Deaf | Female | 16 |
|  | Profoundly Deaf | Female | 7 |
|  | Profoundly Deaf | Male | 13 |
| (B) Users with other types of disabilities | Motor disability | Female | 10 |
|  | Psychosocial disability | Male | 9 |

Finally, as a significant part in this case study, the user experience of a mobile application called Signa App was evaluated. Such an app is designed to support the learning of SignWriting notation. This is a system that uses visual symbols to represent hand shapes, movements, and facial expressions in Sign Languages [17], which makes it perfect to be introduced into the Mexican Sign Language teaching curriculum. The app allows the creation of notes using such notation. Since both the writing notation and the app had never been used before for the participants, it was considered an excellent setting to test UX evaluation methods.

### 4.1    User profile

It is crucial to identify the specific characteristics of each user, as they present different intellectual and academic limitations and abilities. Their level of competence in Mexican Sign Language (LSM) was assessed based on a prior evaluation conducted by their teachers, which classified their proficiency into four levels: insufficient, sufficient, intermediate, and advanced.

The aforementioned data was obtained through an evaluation rubric specifically designed to collect the necessary information. The criteria established in the evaluation rubric were based on four main aspects: a) Comprehension, attitude, and participation; b) Expression, fluency, conversation, and dactylology; c) Structuring, grammatical aspects, and use of classifiers; and d) Professional identity and ethics, along with the learning of alternative and augmentative communication systems in Mexican Sign Language. This classification provided a more accurate overview of each user's skills, facilitating the adaptation of the evaluation process to their specific needs. In **Table 2** we provided demographic information about the participants.

Table 2. User profile for Deaf students.

| Group  | Disability                        | Age | Grade level            | Sign Language level |
|--------|-----------------------------------|-----|------------------------|---------------------|
| User 1 | Profound bilateral hearing loss.  | 16  | 9$^{th}$ grade         | advanced            |
| User 2 | Profound bilateral hearing loss.  | 7   | 2$^{nd}$ grade         | intermediate        |
| User 3 | Profound bilateral hearing loss   | 13  | 4$^{rd}$ grade         | insufficient        |

### 4.2    Description of the sessions

The tests with Deaf users were conducted through an 8-session workshop held at the school, scheduled over four weeks with two sessions per week. This structure ensured consistent participation and allowed time to address emerging questions or challenges.

Session 1: Introduced SignWriting (SW) through classroom presentations, color-coded gloves, illustrative drawings, and explanations linking sign language to SW pictograms.

Session 2: Involved hands-on activities with painted hands and drawings to understand notation, including replicating and transcribing specific signs, and identifying the colors used in the notation.

Session 3: Children practiced writing the notation alphabet using a teaching chart and board activities.

Session 4: Direct observation: Children explored the mobile app while their interactions and reactions were observed.

Session 5: Children played a memory game matching animals with their SW notations.

Session 6: Cognitive walkthrough was applied. Children used the app to sign "Hello" while time, interactions, and errors were recorded.

Session 7: Picture card through prototyping: Children recreated the app on paper using cutouts and drawings based on their imagination.

Session 8 EMODIANA: Children explored the app freely and then selected their mood using a visual emotion board.

The next section analyzes the results of UX methods with Deaf participants and evaluates the accessibility and suitability of the instruments used.

## 5 Results

As mentioned in previous sections, five evaluation methods were chosen to analyze the experience of deaf users when interacting with the application. The results obtained were varied, with both positive and negative aspects. This testing process confirmed that some UX evaluation methods are not effective, as they do not meet the specific needs of Deaf individuals.

### 5.1 User workshop

A user workshop related to the notation was considered, as it is a topic unfamiliar to the participants in the classroom. The children in Groups A and B learned SW concepts such as the alphabet, animals, days of the week, months, and school supplies, while integrating related images in the classroom to support their learning and familiarize themselves with the application, with the teacher's assistance. By the end of the course, students were expected to recognize how the notation enhances classroom learning and communication through the app.

This UX evaluation method allowed a group of potential users (Group A and Group B) to be gathered in the same space, where they were then subjected to a series of tests. Furthermore, the user workshop was effective because it was possible to combine it with the other evaluation methods: a) Cognitive walkthrough, b) Direct observation, c) Picture card, and d) EMODIANA. In the following sections each method is described.

### 5.2    Direct observation

The execution allowed for a detailed observation of how users interacted with the interface, identified potential difficulties, and gathered key information about their experience, making it crucial for ensuring the accuracy and depth of the evaluation results.

Five mobile devices with the Signa App installed were required for the activity, along with a camera to capture images as evidence of the evaluation's development and results. Evaluators organized by observing users, taking notes, and photographing the process.

A recall activity was conducted prior to the test. Then, with the help of the mediator, they were instructed to use an application to generate pictograms in the mobile app. Finally, they were asked to explore the application freely, navigating through all the interfaces and selecting all available options, see **Fig. 1**.

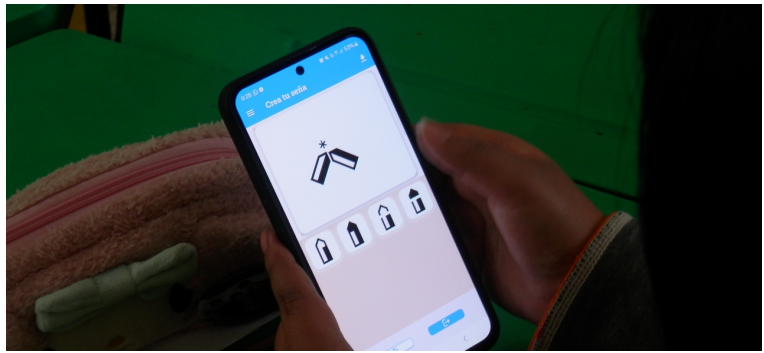

**Fig. 1.** Evaluation evidence: Direct observation.

The direct observation method proved effective, as it did not require significant intervention from the evaluators while users interacted with the software. This approach allowed for a more natural and fluid assessment, where participants were able to engage independently, providing evaluators with a more authentic view of their interaction with the system. However, several important considerations were identified that should be taken into account when applying the method.

### 5.3    Cognitive walkthrough

This evaluation methodology focuses on analyzing how users perceive, understand, and make decisions during their interaction with the interface. It also allowed for the identification of the difficulties experienced by Groups A and B while navigating the application's interfaces. In this way, it facilitated the observation and analysis of how users interact with the interfaces, providing a deep understanding of the mental processes involved in navigation.

Prior to conducting the test, an introductory activity on SignWriting was carried out to familiarize the children with the basic concepts of this notation system. Specifically, both groups of users were provided with the mobile application and asked to perform specific tasks in order to assess their interactions, accuracy, and the time spent completing the activities. The main activities for the children consisted of creating pictograms in SW notation using the application. All users in Groups A and B were able to complete the tasks efficiently and without the need for interventions. However, one of the Deaf users experienced confusion during the test.

### 5.4    Picture card through prototyping

The primary objective of this evaluation was to develop a prototype draft that reflected the ideas and needs of users, aiming to identify and analyze new features that could be incorporated into a future version of the application. In terms of UX, this practice facilitated the schematic representation of the interfaces and interaction flows of the product or service from the user's perspective.

Each user received a blank sheet with representations of three smartphones and was asked to sketch the key screens of the Signa App based on its main functionalities. Participants could illustrate any changes they wished to make to the app based on their previous experiences and add cutouts to the paper prototype. After completing their designs, users shared their opinions and suggestions with the group, as shown in **Fig. 2**.

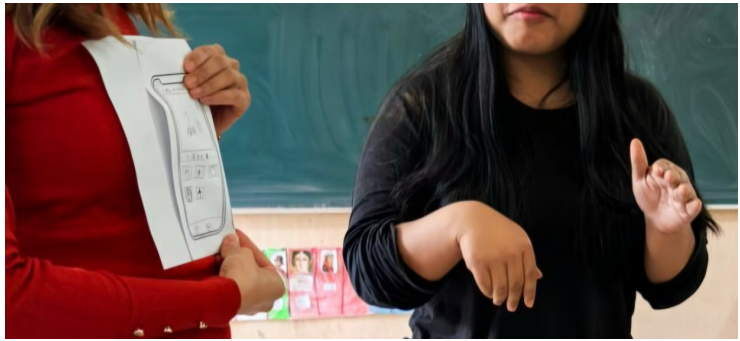

**Fig. 2.** Evaluation evidence: Picture card.

Users' ideas and explanations varied. Group A participants suggested changes to the pictogram creation process, with some finding the interface boring and repetitive. Meanwhile, Group B focused on customization, particularly the creation of avatars to perform written signs through virtual representations. Although all users were able to complete the evaluation, various complications were identified when addressing the activity. These issues are relevant and worth discussing.

### 5.5    EMODIANA

The aim of this evaluation was to identify the emotions experienced by users while using the application. Through this method, called EMODIANA, participants are able to express their emotions using an intensity scale, allowing for an understanding of how they feel during their interaction with the application. This activity was previously discussed with the professor, as in order to carry it out, the children needed to be familiar with the different emotional states presented by the instrument. Emotions have been explored in children's learning through posters placed in the classroom.

To conduct the evaluation, mobile devices were provided to the children, allowing them to interact freely with the application. They were then presented with the EMODIANA board, which displayed 10 emotions represented by an avatar, along with the intensity of each emotion, indicated through colors.

Finally, the evaluation was not successfully completed, as the users were unable to finish the EMODIANA board. Despite being an assessment method designed for children, the tool caused confusion for both Group A and Group B. The accessibility of the method will be a central topic of analysis, as factors were identified that could have influenced the difficulties participants experienced when using the EMODIANA method.

### 5.6   Summary of tasks achieved by method

The implementation of the evaluation methods provided a detailed view of their functionality. **Table 3** summarizes each user's interaction with the different methods applied. In this case, effectiveness is measured by the method completion. This was calculated by the user as the task success rate, and more generally by the method's success rate, considering only totally completed tasks. As observed, User 3 who has a low proficiency in sign language impacts the ability to successfully complete the task.

**Table 3.** Results obtained from UX methods with Deaf students. The values represent if a task was totally completed (2), partially completed (1), and don't completed (0).

| Method | User 1 | User 2 | User 3 | Success Rate (2 values) |
| --- | --- | --- | --- | --- |
| User workshop | 2 | 2 | 2 | 100% |
| Direct Observation | 2 | 2 | 1 | 66.7% |
| Cognitive walkthrough | 2 | 2 | 1 | 66.7% |
| Picture card through prototyping | 2 | 1 | 0 | 33.33% |
| EMODIANA | 1 | 1 | 0 | 0% |
| **Task Success Rate by user** | 90% | 80% | 40% | |

## 6   Discussion

In this section, an analysis will be presented on the accessibility of different methods used for evaluating user experience when applied to deaf individuals. The key challenges and limitations encountered by Deaf participants during testing will be identified and discussed, with a particular focus on aspects that may impede accurate and effective evaluation.

**Limitations.** Although the assessment methods mentioned have been recommended for use with Deaf individuals, their implementation with this specific population revealed several limitations. These difficulties emerged during the evaluations, as it was identified that the linguistic and cultural particularities of deaf individuals were not fully considered, which impacted their effectiveness. The following section will address each of the advantages identified during the trial sessions with Groups A and B.

## 6.1 User workshop

Although this evaluation method offers the advantage of allowing the grouping of a considerable number of users and the possibility of simultaneously executing other evaluation methods, certain deficiencies were identified. Conducting a user workshop involves a series of sessions in which participants acquire new knowledge, with the aim of ensuring they become familiar with and appropriately contextualized before engaging with the application to be evaluated. However, user attendance at the evaluation activities was irregular, despite the implementation of various strategies and resources designed to motivate participation in the tests.

This situation forced the repetition of classes and activities related to SignWriting, which not only extended the time dedicated to these sessions but also caused dissatisfaction among some participants. The need to redo these activities, rather than being seen as an opportunity for reinforcement, was perceived by some as an unnecessary disruption, leading to frustration and discontent within the group. Not only was the originally planned activity schedule affected, but the timeline for the tests also underwent significant modifications. Due to various unforeseen circumstances, many of the scheduled tests had to be repeated to ensure that all users had the opportunity to participate in each of the evaluations.

## 6.2 Direct observation

Both groups proceeded to use the application, allowing for the observation of their interaction with the interfaces. Regarding the Deaf users, each of them made a variety of gestures, signs, and facial expressions to communicate their emotions or thoughts. To properly interpret what occurred, the intervention of the teacher was necessary, as she helped understand the feedback received. In this context, if the observer lacks sufficient knowledge of sign language or is not familiar with the cultural and communicative variations within the deaf community, there is a risk of misinterpreting the observed behaviors, which could affect the accuracy and effectiveness of the evaluation.

One of the deaf participants felt confused during the test, as they were unsure of what to do in the application. As a result, the evaluators' intervention was necessary to guide and motivate the participant to use the application properly. However, this situation led to the user feeling distrustful, which may have influenced their responses, creating potential biases that affected the accuracy of the evaluation. Another important point to consider is that the Deaf community has a unique culture, with its own forms of interaction, values, and norms. Direct observation without an appropriate cultural context may lead to misinterpretations.

### 6.3    Cognitive walkthrough

The explanation of the tasks in Sign Language was not sufficient for Deaf users to fully understand them. This situation caused confusion and delays in the completion of the activities, as the users were unable to effectively grasp all the key aspects of the tasks. Recognizing these challenges, it was decided to employ a different strategy to aid in the users' understanding. For this reason, it was decided to explain the different steps required to complete the task on the whiteboard, using drawings. This approach allowed the users to clearly visualize each action and better understand what was expected of them. The use of visuals not only clarified the instructions but also significantly improved their ability to follow through with the tasks, leading to more efficient and accurate task completion.

One of the Deaf users felt confused while navigating through the interfaces, which caused them to stop during the test and not engage in any further interactions. As a result, the evaluators intervened to assist the user and clarify any doubts that arose during the test. They took the necessary time to carefully demonstrate the different features offered by the application. The evaluators also ensured that the user felt comfortable and understood each step to be followed. However, this intervention may have affected the test results, as it may not have allowed for an authentic and natural interaction from the user, potentially leading to biased data.

### 6.4    Picture card through prototyping

During the prototyping test, multiple challenges were identified for users, especially in Group A. A participant with limited understanding of Mexican Sign Language (LSM) struggled to understand the activity, which led them to replicate the same application design instead of proposing new alternatives. This highlights barriers in the accessibility of evaluation methods, affecting the accuracy and reliability of the results.

Additionally, another user from the same group expressed that the interface seemed boring and suggested improvements. However, due to their limited knowledge of certain signs, they were unable to clearly communicate their ideas, making it difficult to interpret their proposals. These limitations emphasize the importance of adapting evaluation methodologies to ensure a more inclusive experience that accurately represents the needs of Deaf users.

### 6.5    EMODIANA

This method is particularly notable for its application in child populations, as it offers a unique approach to measuring emotions and their intensity in children in a subjective manner. This tool has been widely recognized for its effectiveness in capturing emotional responses, providing valuable insights into the emotional states of users when interacting with a product [4]. However, when the evaluation tool was administered to both deaf and hearing users, it became evident that they faced significant challenges in completing it successfully.

Once both groups completed their interaction with the application, users paid considerable attention to the avatars presented on the EMODIANA board. However, these avatars became a distraction, as each child began to mimic the emotions portrayed by the characters, diverting their focus from the main task. Some participants also took sheets of paper and began drawing each of the avatars, labeling them with the name of the emotion they represented. This behavior suggested that the children were so intrigued by the avatars that they preferred to recreate them manually, which may have further diverted their attention from the main task and hindered the effective application of the evaluation method. As for the colors on the board, which were designed to represent the intensity of the emotion, they went unnoticed during the assessment. The concept of intensity proved difficult for users to understand, despite explanations provided by both the teacher and the evaluators. This difficulty suggests that the relationship between the colors and emotional intensity was not clear or accessible enough for the participants, which impacted the effectiveness of the tool in measuring emotions and their degree of intensity.

## 7      Conclusion

This study analyzes UX evaluation methods from an accessibility perspective, specifically in the context of their application during software testing with Deaf users. The objective of this study was to determine the extent to which the evaluation tools used are appropriate for gathering meaningful information during UX testing with Deaf individuals. Additionally, it aimed to identify the limitations that arise during their application, particularly those related to the communicative, cognitive, and cultural characteristics specific to this population.

The results showed that, although traditional UX methods can be adapted to include Deaf individuals, their application requires substantial modifications that **take into account the linguistic, communicative, and cultural particularities of this community**. Throughout the testing sessions, significant limitations were identified in the implementation of each method, which affected both the quality of the data collected and the participants' overall experience, in summary the challenges are:

- The user workshop, while useful for contextualizing participants, presented logistical and pedagogical challenges related to irregular attendance and the negative perception of content repetition.
- Direct observation, on the other hand, revealed the risk of misinterpretation when observers are not trained in Sign Language and Deaf culture.
- The cognitive walkthrough demonstrated that instructions in LSM do not always guarantee complete understanding, requiring additional visual support (which was provided on a whiteboard).
- In picture cards through prototyping, the language barrier limited the expression of ideas, affecting the quality of the proposals collected.
- EMODIANA showed that while the visual representation of emotions was engaging, it could divert attention from the main task, and the concept of emotional intensity was not fully understood by the users.

Based on the results obtained in this research, it can be concluded that it is essential for the evaluation methods used to undergo specific adaptations. These modifications are necessary to ensure that the evaluations are truly accessible to Deaf individuals, respecting their characteristics and communication skills. The adaptations will not only facilitate the inclusion of this group in evaluation processes but also ensure that the results accurately reflect the experiences and needs of Deaf users.

In conclusion, these findings highlight the need to design and adapt UX evaluation methods that are not only linguistically accessible but also culturally appropriate for the Deaf community. It is essential to involve Deaf professionals and accessibility specialists from the early stages of the methodological design to ensure that the evaluations more accurately reflect the experiences, perspectives, and realities of this population. This research not only sheds light on the existing challenges in this field but also provides a valuable starting point for the development of more inclusive evaluation methods in the UX domain. Furthermore, it emphasizes the need to explore the creation of new mechanisms to facilitate UX evaluation, thereby expanding the possibilities for improving the accessibility of evaluative tools.

**Acknowledgments.** This work is supported by CONAHCYT through the scholarships number CVU: 1084255 and 1299810. Correspondence concerning this article should be addressed to José-Rafael Rojano-Cáceres (rrojano@uv.mx).

**Disclosure of Interests.** The authors have no competing interests to declare that are relevant to the content of this article.